# Detrimental $2p$–$3d$ Hybridisation in Ni Nanosheets Supported on Strontium Dioxide for Catalytic H$_2$ Production, Necessitating Thickness Optimisation

Kabir S. Suraj,[1,2] M. Hussein N. Assadi[2,*]

[1] African University of Science and Technology, Galadimawa, Abuja, Nigeria.
[2] RIKEN Center for Emergent Matter Science, Wako, Saitama 351–0198 Japan
(h.assadi.2008@ieee.org)

We employ accurate density functional theory calculations to examine the electronic structure of three Ni/SrO$_2$ nanostructures containing single-layer, bilayer and four-layer Ni nanosheets. The single Ni layer interacts strongly with the topmost oxygen layer at the Ni/SrO$_2$ interface, resulting in significant surface reconstruction and strong hybridisation between the O $2p$ and Ni $3d$ states. For the bilayer Ni, the layer facing the interface also strongly interacts with the O. However, the second layer retains its geometry. For the four-layer system, none of the Ni layers interacted strongly with O. According to the electronic population analysis, in the thinnest nanosheets, the strong hybridisation with oxygen pulls Ni's $3d$ states away from the Fermi level deeper into the valence band. In these cases, Ni's electronic population that is labile for catalysis in the vicinity of the Fermi level was as little as half of the bilayer nanosheet. Such reduction in the labile $3d$ population has a detrimental effect on Ni's catalytic performance in de-hydrogenating formic acid. Our results demonstrate that there is an optimum dimension for Ni nanoparticles below or above which the catalytic performance deteriorates. Consequently, reducing the Ni dimension to maximise the surface in the hope of better catalytic yield might not be the best strategy as detrimental $p$-$d$ hybridisation takes hold. Smaller might not always be better after all!

## NOMENCLATURE

| *Abbreviations* | |
|---|---|
| DFT | Density Functional Theory |
| DOS | Density of States |
| $E_{\text{Fermi}}$ | Fermi Energy |
| HER | Hydrogen Evolution Reaction |
| OER | Oxygen Evolution Reaction |
| $q$ | Electronic population |

## 1. INTRODUCTION

The science of $3d$ transition metals in catalytic hydrogen production is a critical area of research at the intersection of chemistry, materials science, and renewable energy technology [1]. These metals, which occupy the central portion of the periodic table, have unique electronic structures and versatile catalytic properties that make them indispensable in the quest for efficient and sustainable hydrogen production [2].

One key mechanism for hydrogen production involves the electrocatalytic water-splitting process, where water molecules are split into hydrogen and oxygen using electricity [3]. $3d$ transition metals, such as iron (Fe), cobalt (Co), and nickel (Ni), are at the forefront of this research due to their affordability and abundance when compared to the $4d$ elements. These metals can serve as catalysts at the electrode surface, accelerating the otherwise slow kinetics of water-splitting reactions [4].

Their effectiveness lies in their ability to undergo multiple oxidation states during catalysis, facilitating electron transfer into catalysed molecules' antibonding states in both the hydrogen evolution reaction (HER) and the oxygen evolution reaction (OER) [5, 6]. Additionally, $3d$ transition metals can form alloys or compounds with other materials, enhancing their catalytic activity and stability [7]. Researchers are assiduously investigating various strategies to optimise $3d$ transition metal catalysts. These include alloying with other metals to tune electronic properties, creating nanostructured catalysts to increase surface area, and developing advanced supports or coatings to improve durability [8].

Here, we investigate how the thickness of a transition metal nanosheet on an oxide substrate, nickel and strontium dioxide, in our case, affects the catalytic performance of the system for catalytic hydrogen production. Based on the *d*-band theory for catalysis, the centre of the $d$ states in a transition metal's valence band determines the catalytic potency [9]. In catalytic reactions that extract hydrogen from larger molecules, the electrons from these $d$ states populate the incoming molecules' antibonding states. The molecular antibonding states are empty in stable molecules like water or formic acid. After interacting with the catalyst surface, which fills the molecular antibonding orbitals by the catalyst $d$ electrons, the incoming molecules become unstable, therefore releasing their hydrogen. Subsequently, the location of the catalysts' electronic $d$ states with respect to the Fermi level determines the probability of the electron transfer and catalytic reaction rate.

The results presented here offer an early peek into our theoretical work regarding how varying the thickness of nickel nanosheets on $SrO_2$ affects catalytic performance. The models presented here were constructed by cleaving and interfacing nickel (Ni) in its lowest energy face-centred cubic morphology with $SrO_2$. Ni's lattice parameter is 3.524 Å, $SrO_2$ has lattice parameters of $a = b = 3.550$ Å and $c = 6.550$ Å, crystalising in a tetragonal lattice (Space group number 139). As a result, the basal lattice mismatch is only 0.74% when Ni and $SrO_2$ are interfaced along the [001] direction. A negligible lattice mismatch between these two materials guarantees successful synthesis of the interface, as a more considerable lattice mismatch would lead to phase segregation [10].

## 2. COMPUTATIONAL SETTINGS

Here, we simulated three $Ni/SrO_2$ interface configurations, shown in Fig. 1 through Fig. 3. The common $SrO_2$ support in all configurations comprised of a sheet containing 5 Sr and 10 O atoms and constructed to be symmetric and stoichiometric. On both ends, this support terminated with O. What differentiates these configurations is the thickness of the Ni layer that was positioned on one end of the $SrO_2$ support. Configuration 1 contained one Ni layer, configuration 2 two Ni layers, and configuration 3 contained 4 Ni Layers. Ultimately, to guarantee the realistic simulation of the catalysis-related interface, a sufficiently spacious vacuum slab of no less than 20 Å was employed in all configurations.

We used the VASP package to conduct spin-polarised density functional theory (DFT) calculations within the projector-augmented-wave method [11] and generalised gradient approximation functional [12]. The pseudopotential contained the following electrons: $2s^2 2p^4$ for oxygen, $3d^8 4s^2$ for nickel, and $4s^2 4p^6 5s^2$ for strontium. The energy and force thresholds were set to be $10^{-5}$ eV and 0.01 eV/Å. All other parameters were controlled by setting the precision key to accurate. A dense $13 \times 13 \times 1$ k-point mesh generated using the Monkhorst-Pack scheme and tetrahedron method minimisation with Blöchl corrections were used for Brillouin zone sampling for density of states (DOS) calculations. Geometry optimisation was conducted using the conjugate-gradient algorithm to relax the ions into their instantaneous ground state. Throughout the geometry optimisation procedure, no symmetry constraints were applied, ensuring relaxation to the geometry with the lowest energy. Additionally, all atomic internal coordinates and basal lattice vectors were allowed to relax. The adequacy of these settings was previously tested for metal oxide interfaces [13].

## 3. RESULTS AND DISCUSSIONS

We examined three different nickel nanosheets on the strontium dioxide substrate. The first structure shown in Fig. 1 contains a nickel nanosheet made of one atomic layer. The nanosheet in the second and third structures contained two and four atomic layers, respectively, as shown in Fig. 2 and 3. To determine the availability of nickel's $3d$ states for catalytic reaction, we calculated nickel's electronic population at the vicinity of the Fermi level by integrating the $3d$ partial density of states within a finite energy bracket of $-2$ eV $\leq E_{\text{Fermi}} \leq 0$. This energy bracket has been reported to be the most critical reaction [13, 14].

When the Ni nanosheet is only one layer deep, as in Fig. 1, the substrate's oxygen significantly alters nickel's electronic states. Oxygen's effect stems from its strong affinity to nickel, where the topmost oxygen layer from strontium dioxide breaks away from strontium and bonds to nickel. In essence, when nickel is only one layer deep, it behaves more like an oxide than metal. Accordingly, only 5.20 $e$/Ni are confined within the critical energy bracket immediately available for catalytic redox reaction.

When the Ni nanosheet is made of two atomic Ni layers, as in Fig. 2, the Ni layers behave drastically differently from one another. The first Ni layer that is closer to $SrO_2$ support strongly interacts with the topmost oxygens. Similar to the previous case, this interaction results in O bond breakage from the support and attachment to the lower Ni atoms. However, the upper Ni layer does not interact with O as strongly, remaining metallic. Consequently, the electrons in these two layers behave differently as well. For the lower Ni layer, due to interaction with oxygen, some of Ni's $3d$ states are pulled deeper to the valence band, while for the topmost Ni layer, the $3d$ states mostly gravitate towards higher energies, i.e., closer to the Fermi level. Here, the electronic population within 2 eV from the Fermi level is 3.52 $e$/Ni for the lower layer and 6.73 $e$/Ni for the second layer.

When the Ni nanosheet comprises four atomic layers, unlike the previous two cases, none of the Ni layers interacts strongly with the support layer's oxygen. As shown in Fig. 3, the Ni nanosheets relax to a considerable distance of 4.158 Å from the $SrO_2$ support with no bond rearrangement of surface reconstruction. This interface distance was shorter for both the single-layer and bilayer Ni nanosheets at 3.286 Å and 2.852 Å, respectively. The distribution of Ni's $3d$ states over the valence band is, more or less, uniform across both the layers and energy landscape, resulting in a lower and approximately consistent electronic population within the 2 eV energy bracket. Here, the topmost Ni layer has only 4.62 $e$/Ni, smaller than in previous cases.

Finally, we examined the sensitivity of the electronic population analysis with respect to the 2 eV bracket used so far. The choice of this particular energy bracket was initially inspired by the work of Ceder and colleagues, attributing the density of states within the $-1.64\,\text{eV} \leq E \leq E_\text{Fermi}$ to be the most influential in the redox mechanism [15]. However, wider energy brackets, spanning up to 2.5 eV below the Fermi level, have also been utilised in the literature [13, 16]. We reintegrated Ni's $3d$ partial density of states within two wider brackets of $-2.5\,\text{eV} \leq 0 \leq E_\text{Fermi}$ and $-3\,\text{eV} \leq 0 \leq E_\text{Fermi}$. According to the results presented in Table 1, although the absolute electronic population increases for wider energy ranges, the relative $3d$ electronic population per layer varies little, showing a standard deviation below 0.1. However, for broader energy ranges, the difference in layer-resolved electronic population vanishes quickly.

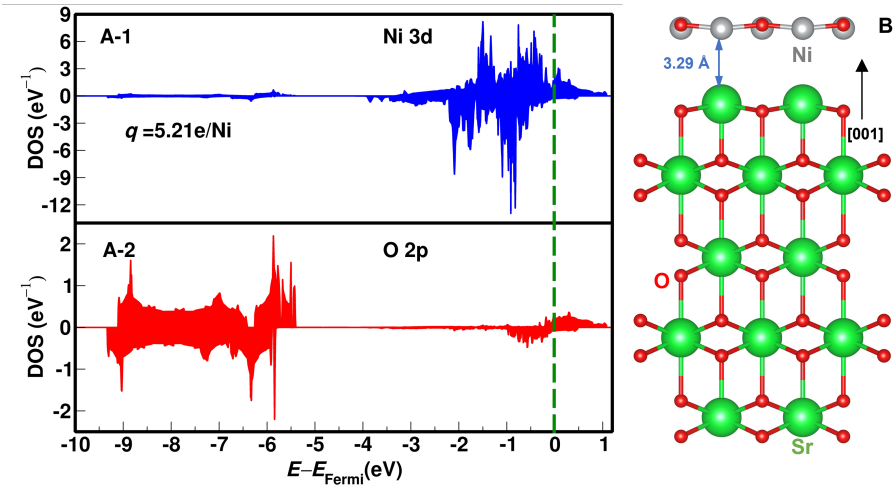

**Fig. 1.** A-1 and A-2) Site-projected partial density of states of the Ni $3d$ and the outermost layer of the O $2p$ states. $q$ values denote the partial electronic population in the Ni $3d$ orbitals whose states lie within $-2.0\,\text{eV} \leq E \leq E_\text{Fermi}$. B) The optimised structure of the strontium dioxide and Ni interface.

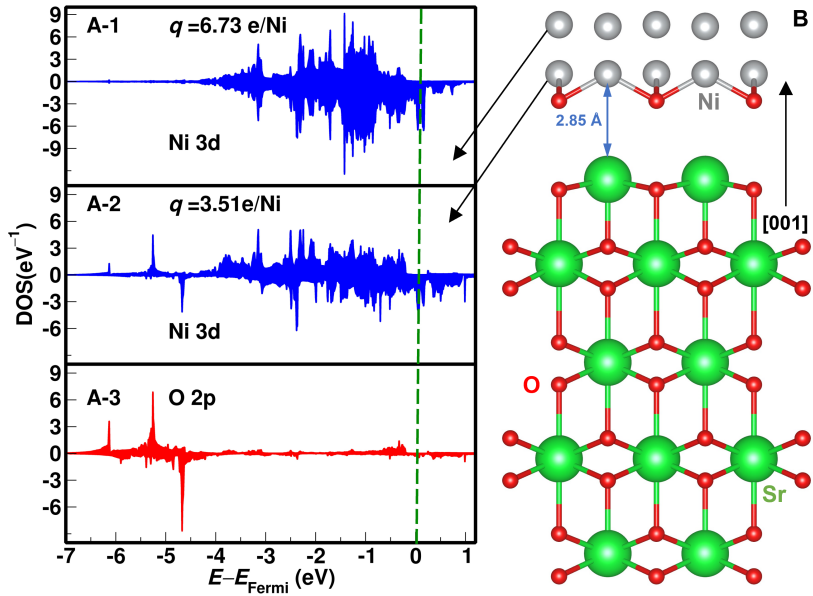

**Fig. 2.** A-1 through A-3) showing the site-projected partial density of states of the Ni $3d$ and the outermost layer of the O $2p$ states resolved by atomic layers. $q$ values denote the partial electronic population in the Ni $3d$ B) The optimised structure of the strontium dioxide and Ni interface.

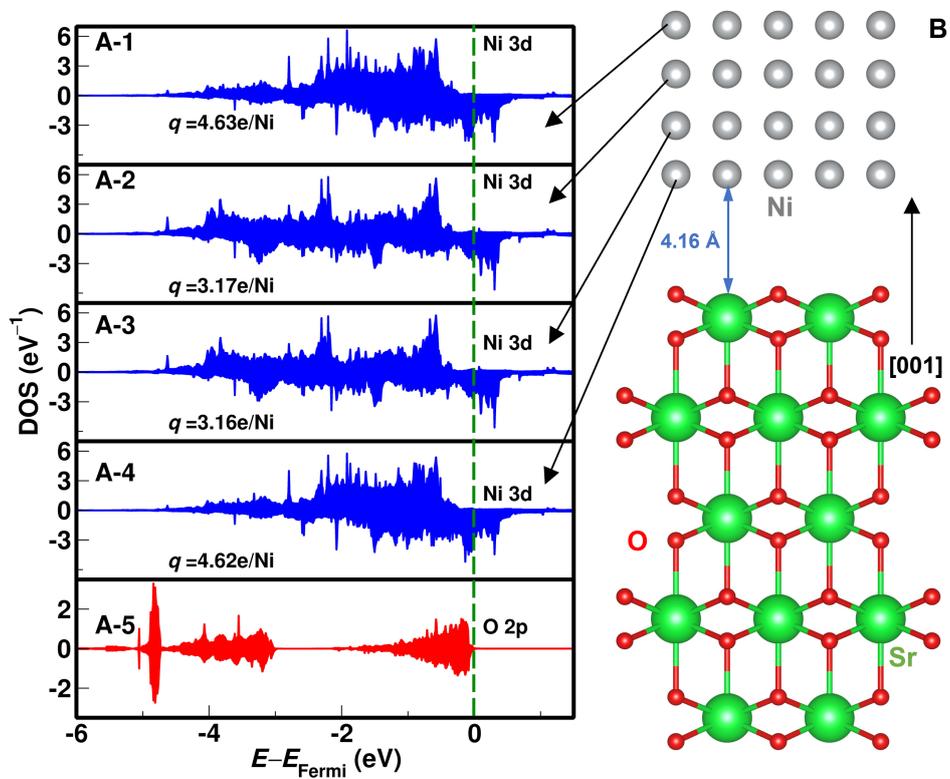

**Fig. 3.** A-1 through A-5) Site-projected partial density of states of the Ni $3d$ and the outermost layer of the O $2p$ states resolved by atomic layers. $q$ values denote the partial electronic population in the Ni $3d$ orbitals whose states are within $-2.0 \leq E \leq E_{\text{Fermi}}$. B) The optimised structure of the strontium dioxide and four-layer Ni interface.

**Table 1.** Partial electronic population in the Ni $3d$ orbitals for different energy levels below the Fermi level.

|  | $-3.0 \leq E \leq E_{\text{Fermi}}$ | $-2.5 \leq E \leq E_{\text{Fermi}}$ | $-2.0 \leq E \leq E_{\text{Fermi}}$ |
|---|---|---|---|
| Single-Layer | 6.17 | 5.89 | 5.20 |
| Bilayer | 8.84 | 8.22 | 6.73 |
|  | 5.04 | 4.45 | 3.51 |
| Four-Layered | 5.82 | 5.39 | 4.62 |
|  | 4.74 | 4.07 | 3.16 |
|  | 4.76 | 4.08 | 3.17 |
|  | 5.83 | 5.39 | 4.61 |

## CONCLUSIONS

In summary, our study reveals how critical the size of Ni nanosheet is in enhancing the catalytic performance of Ni/SrO$_2$ nanostructures used for H$_2$ production. Ni-O $3d$-$2p$ hybridization sets an optimum amount of two Ni layers that can be interfaced with SrO$_2$ emphasizing the experimental need for immaculate control over the synthesis of Ni nanoparticle size for improved catalysis.

## ACKNOWLEDGEMENTS

The computing resources were provided by the HOKUSAI system at RIKEN. This work was also supported by funding obtained from the EIG CONCERT-Japan under grant 2023-MLALH. KSS acknowledges the financial support from the IPA program at Riken, the Marubun grant and the grant from the African development bank.